\documentstyle[aas2pp4,flushrt,epsfig,twoside]{article}
\newcommand{\ee}[1]{\times 10^{#1}} 
\newcommand{\reaction}{{${\rm^9Be}(p,\alpha){\rm^6Li}$}} 
\newcommand{\kB}{k_B}		
\newcommand{\K}{\mbox{$\rm\,K$}} 
\newcommand{\gram}{\mbox{$\rm\,g$}} 
\newcommand{\second}{\mbox{$\rm\,s$}} 
\newcommand{\Myr}{\mbox{$\rm\,Myr$}} 
\newcommand{\Gyr}{\mbox{$\rm\,Gyr$}} 
\newcommand{\fermi}{\mbox{$\rm\,fm$}} 
\newcommand{\ang}{\mbox{$\rm\,\AA$}} 
\newcommand{\cm}{\mbox{$\rm\,cm$}} 
\newcommand{\MeV}{\mbox{$\rm\,MeV$}} 
\newcommand{\keV}{\mbox{$\rm\,keV$}} 
\newcommand{\barn}{\mbox{$\rm\,barns$}} 
\newcommand{\Msun}{\mbox{$M_\odot$}} 
\newcommand{\EG}{E_{\rm G}}	
\newcommand{\Teff}{\mbox{$T_{\rm\!eff}$}} 
\newcommand{\tdepl}{t_{\rm depl}} 
\newcommand{\Hyd}{\mbox{$\rm H$}} 
\newcommand{\He}{\mbox{$\rm^4He$}} 
\newcommand{\sixLi}{\mbox{$\rm^6Li$}} 
\newcommand{\Li}{\mbox{$\rm^7Li$}} 
\newcommand{\Be}{\mbox{$\rm^9Be$}} 
\newcommand{\tenB}{\mbox{$\rm^{10}B$}} 
\newcommand{\B}{\mbox{$\rm^{11}B$}} 
\newcommand{\Fe}{\mbox{$\rm Fe$}} 
\begin{document}
\pagestyle{myheadings}
\markboth{Brown}{Subthreshold Resonance for Stellar Beryllium Depletion}
\title{
   Implications of a Subthreshold Resonance for Stellar Beryllium
   Depletion
}
\author{
   Edward F. Brown
}
\affil{
   Department of Physics and Department of Astronomy\\
   601 Campbell Hall, Mail Code 3411, University of California,
   Berkeley, CA 94720--3411\\
   e-mail: ebrown@astron.berkeley.edu
}
\thispagestyle{empty}
\begin{center}
\large To appear in {\sc the Astrophysical Journal}, 10 March 1998
\end{center}
\vspace*{-20pt}

\begin{abstract}

Abundance measurements of the light elements lithium, beryllium, and
boron are playing an increasingly important role in the study of stellar
physics.  Because these elements are easily destroyed in stars at
temperatures $(2\mbox{--}4)\ee{6}\K$, the abundances in the surface
convective zone are diagnostics of the star's internal workings.
Standard stellar models cannot explain depletion patterns observed in
low mass stars, and so are not accounting for all the relevant physical
processes.  These processes have important implications for stellar
evolution and primordial lithium production in big bang nucleosynthesis.
Because beryllium is destroyed at slightly higher temperatures than
lithium, observations of both light elements can differentiate between
the various proposed depletion mechanisms.  Unfortunately, the reaction
rate for the main destruction channel, \reaction, is uncertain.  A level
in the compound nucleus \tenB\ is only 25.7\keV\ below the reaction's
energetic threshold.  The angular momentum and parity of this level are
not well known; current estimates indicate that the resonance entrance
channel is either $s$- or $d$-wave.  We show that an $s$-wave resonance
can easily increase the reaction rate by an order of magnitude at
temperatures $T\approx4\ee{6}\K$.  Observations of $M<\Msun$ stars can
constrain the strength of the resonance, as can experimental
measurements at laboratory energies lower than 30\keV.

\end{abstract}


\section{Introduction}\label{s:introduction}

In recent years the abundances of the light elements lithium,
beryllium, and boron have been increasingly used as diagnostics of
stellar physics.  The fragility of these elements to proton capture
makes their abundances exquisite probes of a star's internal machinery.
Abundance measurements of both beryllium and lithium can differentiate
between proposed depletion mechanisms.  This presupposes that the
destruction rate is accurately known.  Unfortunately, the reaction cross
section for one of the beryllium destruction channels, \reaction, is
uncertain.  Just $25.7\keV$ below the energetic threshold for the
reaction, there is a level in the compound nucleus \tenB.  Uncertainties
in the angular momentum and parity of the compound nucleus level
prohibit an accurate theoretical calculation of the reaction rate, and
laboratory measurements of the cross-section to date are inconclusive.
Current astrophysical rates (e.g., Caughlan \& Fowler 1988, hereafter
\cite{cau88}) do not account for the cross section's uncertainty.  In
this paper, we describe astrophysical topics for which accurate reaction
rates are needed, we estimate how large a correction to the reaction
rate might be, and discuss how astrophysical observations may be used
to inform the nuclear physics as to the possible magnitude of the
reaction cross section.

In \S~\ref{s:motivation}, we outline the motivation for investigating
the contribution of compound nucleus formation (\tenB) to the
destruction rate for beryllium.  Section \ref{s:FGap} reviews lithium
and beryllium depletion observations in F stars (the ``F Gap'') and
illustrates how abundance ratios of beryllium to lithium can constrain
the various mechanisms for destroying these elements.  We discuss in
\S~\ref{s:LowMassDepletion} the depletion patterns in subsolar-mass
stars.  We list different scenarios for destroying beryllium and lithium
in these stars, and the consequences of each scenario.

After describing the importance of correctly calculating the destruction
rate for beryllium, we then outline (\S~\ref{s:CrossSection}) the
uncertainties in the \reaction\ cross section.  Section
\ref{s:experiment} describes experimental studies of the cross section
and the quantum numbers characterizing the \tenB\ compound nucleus.  We
then recalculate the astrophysical $S$-factor and estimate the thermally
averaged cross section in \S~\ref{s:calculation}.  In
\S~\ref{s:constraints} we discuss the prospects for using
astrophysical observations to constrain the reaction $S$-factor.

\section{Motivation: the Importance of Beryllium Abundances}
\label{s:motivation}

The light element isotopes \sixLi, \Li, \Be, \tenB, and \B\ both trace
Galactic chemical evolution and probe the interior workings of stars.
The high stability of helium nuclei dominates the nuclear physics of
these elements; indeed, all three are easily destroyed by proton capture
at stellar temperatures $T=(2\mbox{--}6)\ee{6}\K$.  Only \Li\ is
produced in appreciable quantities by standard big-bang nucleosynthesis
(BBN).  Inhomogeneous BBN models (\cite{boy89}; \cite{kaj90}) can
produce beryllium, but this would imply a ``plateau'' in beryllium
abundances for very low metallicity stars.  There is no firm
observational evidence for such a plateau; instead, $[\Be]\equiv
12+\log(N_{\rm^9Be}/N_{\rm H})$ appears to be linearly correlated with
$[\Fe/\Hyd]$ for abundances of $[\Be]\gtrsim -1$ (\cite{boe93};
\cite{gil92}; \cite{reb95}; \cite{rya92}).  All three elements are also
produced by spallation reactions, in which high-energy H and He nuclei
bombard CNO elements.  Formation by the reverse reaction, bombardment of
H and He nuclei by CNO elements accelerated in supernovae, may also be
important in star-forming regions (\cite{cas95}).  Because spallation is
the only mechanism for producing \Be\ (there is no low-energy formation
channel), measurements of beryllium trace the cosmic-ray history of the
galaxy (for a review see \cite{ree94}).

Observations of surface beryllium abundances in stars can reveal the
cause of discrepancies between observed surface abundance patterns and
standard stellar models.  A challenge to any stellar theory is to
explain the lithium and beryllium F gap, the solar abundances, and the
increased depletion with decreasing effective temperature for F-, G-,
and K-type stars.  Beryllium observations, when used in conjunction with
lithium observations, provide a powerful diagnostic of the stellar
physics, as we discuss in \S\S~\ref{s:FGap} and
\ref{s:LowMassDepletion}.

Of these three elements, lithium is most easily observed, because of the
Li\/I optical transition at $\lambda=6708\ang$.  In contrast, the
abundance of beryllium is inferred from measurement of the Be\/II
doublet in the near-UV ($\lambda=3130\ang$ and 3131\ang).  Although
Be\/I has a transition at $\lambda2349$, observable with the {\em Hubble
Space Telescope\/}, the line strength depends only weakly on the
abundance (\cite{gar96}).  Only recently have UV observations attained
sufficiently high signal-to-noise ratios to allow spectral analysis of
this doublet for low-mass stars (the spectra of solar-mass stars are
typically crowded about this wavelength).  Very low mass stars
($M\lesssim0.3\Msun$) are quite UV dim, so they offer little prospect
for beryllium observations in the near future.  This is unfortunate, as
beryllium observations in these very low mass stars can easily constrain
the uncertain reaction rate (see \S~\ref{s:constraints}).

\subsection{The Lithium and Beryllium F Gap}\label{s:FGap}

Severe lithium depletion in Hyades F stars was first observed by
Boesgaard \& Tripicco (1986).  This gap is remarkable: the lithium
abundance falls by a factor of 100 over a narrow ($\sim600\K$) range
of effective temperatures about $\Teff\sim 6600\K$.  Standard stellar
models do not predict this feature, and so do not account for all the
relevant physics.  Several mechanisms have been proposed to explain the
removal of light elements from the surfaces of these stars.  Among those
which do not rely on nuclear burning are mass loss (\cite{swe92}) and
microscopic diffusion (\cite{ric93}).

Mass loss can nominally explain the Hyades F gap, although it requires a
finely tuned mass-loss rate; stars in the gap must have higher mass-loss
rates than stars of slightly hotter and cooler \Teff.  Because the
lithium preservation region is more shallow than the beryllium
preservation region, mass loss would remove essentially all of the
surface lithium before diluting the beryllium-rich layers.  If the
beryllium destruction rate were greatly enhanced, then a star could
exist that was slightly depleted in beryllium and yet had some lithium
remaining.

In contrast, microscopic diffusion predicts a beryllium gap in the
Hyades (\cite{ric93}).  At an age of 700\Myr, the Be gap will be
centered about 70\K\ cooler than the lithium gap.  Because other
species, most notably \He, will also sink relative to hydrogen,
increasing the diffusivity has strong consequences for inferred stellar
properties.  Metal abundances in both lithium- and beryllium-depleted
F stars (\cite{boe86a}) do not show any obvious trend, which disfavors
pure diffusion models.  One strong prediction of the diffusion models,
which also include radiative levitation, is the presence of a lithium
``bump,'' a range of effective temperatures ($6900\K<\Teff<7100\K$) for
which stars are overabundant.  A corresponding Be bump will also exist
for $6700\K<\Teff<6900\K$ (\cite{ric93}).

Another class of mechanisms slowly mix lithium, beryllium, and boron to
depths where reactions can occur.  The abundance ratio of these three
elements is then sensitive to their proton-capture rates, which set the
depth at which destruction of each species occurs.  This mixing can be
caused by, for example, internal waves (\cite{mon96}), meridional
circulation (\cite{zah92}; Chaboyer, Demarque, \& Pinsonneault 1995a,
1995b; Charbonnel, Vauclair, \& Zahn 1992), and rotation-induced
turbulence (\cite{zah92}; \cite{cha94b}).  Horizontal turbulence
inhibits chemical advection so that angular momentum transport is much
more efficient than chemical species transport; the movement of species
is then a diffusive process (\cite{cha92b}).  The F gap is explained if
angular momentum loss via a stellar wind drives the circulation.  Stars
on the hot side of the lithium gap do not spin down during their
main-sequence lifetime, but stars on the cool side do (\cite{kra67}).

While the morphology of the lithium gap alone can differentiate between
depletion models (e.g., \cite{bal95}), each proposed mechanism also
affects the abundances of beryllium and boron.  Deliyannis \&
Pinsonneault (1997) discuss how observations of both lithium and
beryllium can be used to explore the nature of the nonstandard physics
responsible for the lithium F gap.  In particular, 110~Her, which is
depleted in lithium by a factor of 100--200 and beryllium by a factor of
about 10 (\cite{boe86a}) is possibly an F-type star caught ``in the
act'' of depleting both elements.  Recent boron measurements, using the
{\em Hubble Space Telescope\/}, find no boron deficiencies in Li- and
Be-depleted F stars (\cite{boe98}).  As boron burns at even higher
temperatures (i.e., at greater depths) than beryllium and lithium, these
observations argue in favor of some type of mixing as the means of
destroying lithium and beryllium.

\subsection{Depletion for $M\lesssim\Msun$}\label{s:LowMassDepletion}

For effective temperatures below $\sim6000\K$, the surface convective
zone deepens, and not unexpectedly the abundance of lithium (and
possibly beryllium) decreases with declining \Teff.  Observations show
that lithium is depleted in the sun by a factor $\sim100$ relative to
meteoric values ($[{\rm Li}]=3.31\pm0.04$; \cite{and89}).  Recent
measurements (\cite{kin97}) suggest that beryllium is also depleted,
relative to the meteoric abundance ($[\Be]=1.42\pm0.04$; \cite{and89}),
by a factor of 1.4--3.0 in the sun.  This depletion is mirrored in the
solar-like stars $\alpha$~Cen~A and $\alpha$~Cen~B (\cite{kin97};
\cite{pri97}).

Comparisons between Pleiades (age $\sim 100\Myr$), Hyades (age
$\sim600\Myr$), and M67 stars (age $\sim5\Gyr$) (see Fig.~1 in
\cite{rya95}) indicate that G and K dwarfs deplete lithium on the main
sequence.  Stars with effective temperatures $\Teff\lesssim6000\K$ show
an increasing depletion of lithium with decreasing \Teff.  Although
standard models with convective overshoot can match the Pleiades'
abundance pattern, they are unable to duplicate the Hyades' depletion.
As with F-gap stars, a variety of different mechanisms exist to explain
this abundance pattern.

Although mass loss is somewhat consistent with the formation of a
lithium F gap, it cannot explain the decrease in lithium abundance with
declining \Teff\ for Hyades G dwarfs (\cite{swe92}).  Diffusion appears
necessary for agreement between standard solar models and helioseismic
constraints on the depth of the convective zone (Bahcall, Pinsonneault,
\& Wasserburg 1995) and the sound speed (\cite{bah97}).  Chaboyer,
Demarque, \& Pinsonneault (1995a, 1995b) considered both rotational
mixing and microscopic diffusion (for $M<1.3\Msun$).  They found that
although microscopic diffusion alone was insufficient to match
simultaneously observations of different clusters, a model incorporating
both rotation and diffusion was.  Observations of lithium abundances in
halo stars also do not agree with pure diffusion models (\cite{cha94a}).
If depletion is due to meridional circulation, then lithium should be
better preserved in tidally locked binaries (\cite{zah94}); there is
some evidence for this (\cite{rya95}).

The different scenarios for destroying lithium have cosmological
consequences.  Models that use internal mixing to destroy lithium and
beryllium imply that lithium is not as well preserved in old cluster
members as previously thought (\cite{cha94a}).  Differences in the
lithium abundances of identical M92 subgiants might imply differential
depletion due to rotation-induced mixing (Deliyannis, Boesgaard, \& King
1995), and either a higher primordial lithium abundance or some form of
chemical enrichment.  An enhanced primordial lithium abundance is a
challenge to standard BBN (see, for example, Fig.\ 13 of \cite{wal91}).
Diffusion of \He\ from the surface lowers \Teff\ (\cite{ric93}) and so
reduces the ages of halo dwarfs by 2--3\Gyr\ (\cite{cha92a};
\cite{cha94a}).  Diffusion of heavy elements mitigates this effect, but
an age reduction of $\sim1\Gyr$ appears unavoidable (\cite{cas97}).

Stars of mass $M\lesssim\Msun$ deplete lithium and beryllium while
contracting to their main sequence radius.  Stars less massive than
$\sim0.5\Msun$ (the exact mass depends on the isotope) destroy that
element before developing a radiative core (e.g., \cite{ush98});
otherwise, destruction of a given light element occurs at the base of
the outer convective zone.  As the convective zone moves outward (in
mass coordinates), the temperature at the base of the convective zone
passes through a maximum, which is typically hot enough for lithium, and
to a lesser extent beryllium, to burn.  As a result, zero-age
main-sequence (ZAMS), subsolar-mass stars display abundances that
reveal the history of their convective zone's base.  These stars offer
the best astrophysical environment for examining the nuclear physics
(see \S~\ref{s:constraints}).

\section{The \mbox{\protect\boldmath\reaction} Cross-Section}
\label{s:CrossSection}

An important part of the microphysics in any stellar model is the
destruction rates of the light elements.  The lithium rates have
recently been adjusted (\cite{rai93}) slightly.  For the case of
beryllium, there is a potentially strong correction to the \cite{cau88}
rates.  Just 25.7\keV\ below the energetic threshold (see Table
\ref{t:levels}) of the reaction \reaction\ lies an excited level
($E_{\rm exc} = 6.56\MeV$) of \tenB\ (\cite{ajz88}).  This state
primarily decays by $\alpha$-emission, and so forms a resonant channel
for the \reaction\ reaction.  As we discuss in \S~\ref{s:experiment},
the $S$-factor used by CF88 is based on the resonant level having an
angular momentum and parity $J^\pi=2^+$.  Current knowledge (see
\cite{ajz88}) is that the angular momentum and parity of the \tenB\
state is either $J^\pi=2^-$ or $J^\pi=4^-$, with $4^-$ thought the
better guess.  The angular momentum and parity of \Be\ is $(3/2)^-$.
Hence, if the resonant state has $J^\pi=2^-$, the reaction can proceed
with $\ell=0$; if instead $J^\pi=4^-$, the lowest entrance channel
available is $\ell=2$.  Because the strength of the resonance strongly
depends on the angular momentum, the reaction rate can vary by several
orders of magnitude for these alternatives.  We estimate the cross
section in \S~\ref{s:calculation}.

\subsection{Experimental Measurements}
\label{s:experiment}

The reaction $\Be+p$ was of interest historically for fusion power
generation because of its large cross section at low energies.  Sierk \&
Tombrello (1973) studied the reactions $\Be(p,d)2\,\He$ and
\reaction\ at center-of-mass (CM) energies of 30--630\keV.  They
concluded that the 310\keV\ resonance (see Table \ref{t:levels}) had
negative parity because of $s$-wave proton formation.  Because of
low-energy asymmetries in the $d$ and $\alpha$ angular distributions, they
then assigned a positive parity to the $E_{\rm exc}=6.56\MeV$ level.  In
their fit to the reaction cross-section, they used $J^\pi=2^+$, so that
the entrance channel was p-wave.  Mindful of the astrophysical
importance of this reaction, they estimated the $S$-factor for both
\reaction\ and $\Be(p,d)2\,\He$ to be $S(0)=35^{+45}_{-15}\MeV\barn$.
The contribution of the resonance to the total $S$-factor was between 20\%
and 40\%.  Uncertainties in the compound nucleus formation allow the
possibility that the $S$-factor could actually increase at lower energies.

Measurements of the angular momentum and parity of the levels in \tenB\
have also been performed via $\B(^3He,\alpha)\tenB(\alpha_0)\sixLi$,
$\sixLi(\alpha,\alpha)\sixLi$, $\Be(^3He,d)\tenB$, and $\Be(d,n)\tenB$
reactions.  Young, Lindgren, \& Reichart (1971) used correlations in
$\B(^3He,\alpha)\tenB(\alpha_0)\sixLi$ to determine that $J(E_{\rm
exc}=6.56\MeV)\ge 3$.  Interference between this level and the one at
$E_{\rm exc}=7.00\MeV$, which was presumed to have $J^\pi=3^+$ [note
that the currently accepted value is $(1,2)^+$; Table \ref{t:levels}],
led them to infer that $J^\pi(E_{\rm exc}=6.56\MeV) =3^-$ or $4^-$.
Measurements from $\sixLi(\alpha,\alpha)\sixLi$ scattering
(\cite{bal71}) suggested that $J^\pi$ equaled either $4^-$ or $2^-$,
with $2^-$ preferred.  Park, Niiler, \& Lindgren (1971), using the
reaction $\Be(d,n)\tenB$ at laboratory energies in excess of 7\MeV,
assigned $J^\pi=3^-$ to the 6.13\MeV\ level and a negative parity to the
$6.56\MeV$ level.  Bland \& Fortune (1980) found that a {\em positive\/}
parity for the 6.13 and 6.56\MeV\ levels gave a better fit to angular
distributions in $\Be(^3He,d)\tenB$.  However, the spectroscopic
strengths were then much larger than expected.  An assignment of
$J^\pi=3^-$ to the 6.13\MeV\ level agreed well with predicted strengths;
because the angular distribution of the 6.56\MeV\ level was very similar
to that of the 6.13\MeV\ level, Bland \& Fortune (1980) concluded that
both levels have negative parity.  This choice also gave the greatest
consistency between the $(d,n)$ and $(^3He,d)$ reactions.  Their
conclusion was tempered by the poor angular distribution fit of their
distorted-wave Born approximation, which they were unable to explain.
Recent shell-model calculations (\cite{war92}) favor $J^\pi = 4^-$.
Recently, Zahnow et al.~(1997) measured $S(E)$ down to laboratory
energies of 16\keV, which is significantly less than Sierk \&
Tombrello's (1973) measurement ($E_{\rm lab}<28\keV$).  Zahnow et
al.~(1997) found that $S(E)$ increases sharply with decreasing proton
energy, and that $S(E_{\rm lab}=15.93\keV)=51\pm16\MeV\barn$, which is
greater than the previously measured $S$-factor,
$S(0)=35^{+45}_{-15}\MeV\barn$ (Sierk \& Tombrello 1973).

\begin{deluxetable}{rrrrr}
\large\renewcommand{\baselinestretch}{0.7}\small
\tablewidth{0pt}
\tablecaption{Levels in $\tenB$ near the \reaction\ reaction threshold
($E=6.5857\MeV$).
\label{t:levels}}
\tablehead{
   \colhead{Level energy} & \colhead{Resonance energy} &
   \colhead{Angular Momentum,} & \colhead{Level
   width} & \colhead{Decay}\\
   \colhead{(MeV)} & \colhead{(keV)} & \colhead{Parity\tablenotemark{a}}
   & \colhead{(keV)} & \colhead{Channel}
}
\startdata
6.127 & -459\phd\phn & $3^-$ & $2.36\pm0.03$ & $\alpha$ \nl
6.560 & -25.7 & $(4)^-$ & $25.1\pm1.1$\phn & $\alpha$ \nl
6.873 & 287\phd\phn & $1^-$ & $120\pm5$\phn\phn\phd & $\gamma$,p,d,$\alpha$ \nl
7.002 & 416\phd\phn & $(1,2)^+$ & $100\pm10$\phn\phd & p,d,$\alpha$ \nl
\enddata
\tablecomments{The resonance energy is measured with respect to the
energetic threshold for \reaction.}
\tablerefs{\protect\cite{ajz88}}
\tablenotetext{a}{Parentheses indicate that the values are uncertain.}
\end{deluxetable}

It is clear that the parameters of the \tenB\ level $E_{\rm
exc}=6.56\MeV$ are still uncertain.  We note that the $S$-factor used by
CF88 is the same as that measured by Sierk \& Tombrello (1973), which
assumed a {\em positive\/} parity (and hence a p-wave entrance channel
for \reaction) for the \tenB\ level.  Given the importance of beryllium
abundance observations to astrophysics, accurate measurements of the
cross-section at low energies are greatly needed.  For the purposes of
this paper, we treat both possibilities ($J^\pi=4^-$ or $2^-$) as
equally possible, and recalculate the $S$-factor and reaction rate for
each case.

\subsection{An Estimate of the Resonant Cross-Section and Reaction Rate}
\label{s:calculation}

For a reaction that proceeds via a subthreshold compound nucleus, the
cross-section is given by the Breit-Wigner single level formula,
\begin{eqnarray}\label{e:cross-section}
\lefteqn{\sigma(p,\alpha) = \left(\frac{\pi}{k^2}\right)}\\ \nonumber
  && \times
      \left[\frac{2J_{\rm B}+1}{(2J_p+1) (2J_{\rm Be}+1)}\right]
      \left[\frac{\Gamma_p\Gamma_\alpha}{(E-E_r)^2 +
      (\Gamma/2)^2}\right],
\end{eqnarray}
where the first term is the geometrical cross-section ($k$ is the
wavevector), and the second is the statistical factor.  The total width
of the state, at an energy $E_r$ relative to the energetic threshold, is
$\Gamma$, which for \reaction\ is just the width of the $\alpha$-channel
(\cite{ajz88}), and $\Gamma=\Gamma_{\alpha}=25.1\keV$.  The
cross-section is therefore completely determined once the proton width
$\Gamma_p$ is known.

We parameterize the entrance channel $\Gamma_p$ by the
dimensionless reduced width $\theta_\ell^2$,
\begin{equation}\label{e:Gamma_p}
   \Gamma_p(\ell;E) = \frac{3\hbar v}{R}P_\ell^2 \theta_\ell^2,
\end{equation}
where $R$ is the strong-force interaction radius (we use $R=3.88\fermi$),
$v$ is the relative velocity in the CM frame, and $P_\ell$ is the
penetration factor for proton angular momentum $\ell$.  In terms of the
Coulomb wave functions $F_\ell$ and $G_\ell$, the penetration factor is
$P_\ell = [F_\ell(E;R)^2 + G_\ell(E;R)^2]^{-1}$.  We use the low-energy
approximation for $F_\ell$ and $G_\ell$ (\cite{abr65}),
\begin{mathletters}
\begin{eqnarray}
   F_\ell\!\! &\approx&\!\! \frac{(2\ell+1)!C_\ell(\eta)}{(2\eta)^{\ell+1}}
      (2\eta\varrho)^{1/2}
      I_{2\ell+1}\left[2\left(2\eta\varrho\right)^{1/2}\right]\\ 
   G_\ell \!\!&\approx&\!\! \frac{2(2\eta)^\ell}{(2\ell+1)!C_\ell(\eta)}
      (2\eta\varrho)^{1/2}
      K_{2\ell+1}\left[2\left(2\eta\varrho\right)^{1/2}\right],
\end{eqnarray}
\end{mathletters}
where
\begin{eqnarray}
   2\pi\eta &\equiv& \left(\frac{\EG}{E}\right)^{1/2}\\ \nonumber
      &\equiv&
      \left[\left(\frac{2\pi Z_{\rm Be}Z_{\rm H} e^2}{\hbar}\right)^2
      \left(\frac{\mu}{2}\right)\right]^{1/2} E^{-1/2}
\end{eqnarray}
parameterizes tunneling through the Coulomb barrier ($\EG$ is the Gamow
energy, $Z_{\rm Be}$ and $Z_{\rm H}$ are respectively the charge numbers
of the \Be\ and H nuclei, and $\mu=0.9 m_u$ is the reduced mass of the
$p+\Be$ system), $\varrho = (2\mu R^2E/\hbar^2)^{1/2}$ accounts for
the centrifugal barrier, $I_{2\ell+1}$ and $K_{2\ell+1}$ are the
modified Bessel functions of order $2\ell+1$, and
\begin{equation}
   C_\ell(\eta) =
      2^\ell\exp\!\left(-\frac{\pi\eta}{2}\right)
      \frac{|(\ell+i\eta)!|}{(2\ell+1)!}.
\end{equation}
Numerically, $2\pi\eta = 118E^{1/2}$ and $\varrho = 0.0255 E^{1/2}$,
when $E$ is measured in \keV.  At typical astrophysical energies,
$E\sim10\keV$, the penetration factor reduces to the standard WKB form
(e.g., \cite{cla83}) and is dominated by the term
$\exp[-(\EG/E)^{1/2}]$.

We qualitatively estimate $\theta_0^2$ by approximating the nuclear
potential as a square well (see, e.g., \cite{bla79}).  If we construct a
wave packet out of states of average level spacing $D$, then the period
of oscillation is $D/2\pi\hbar$.  Multiplying this by the transmission
probability $4k/K$, where $K$ and $k$ are respectively the wavevectors
inside and outside the potential well, and by the Coulomb barrier
penetration probability $P_0$, we obtain an estimate of the lifetime of
the state, which is just $\Gamma_p/\hbar$.  Equating this
estimate with the equation for $\Gamma_p$, equation
(\ref{e:Gamma_p}), we then obtain an estimate for $\theta_0^2$,
\begin{equation}\label{e:estimate}
   \theta_0^2 \approx \frac{2}{3\pi}
   \left(\frac{\mu}{2E_{\rm exc}}\right)^{1/2} 
   \frac{RD}{\hbar},
\end{equation}
where $E_{\rm exc}=6.56\MeV$ is the energy of the resonant level.  We
consider two values of $D$.  First, the closest level with $J^\pi=2^-$
is 310\keV\ above $E_{\rm exc}$; using this in equation
(\ref{e:estimate}) implies that $\theta_0^2 = 0.011$.  If instead we
average the energy separations of the two nearest levels with
$J^\pi=2^-$, we obtain $D=880\keV$, which implies that $\theta_0^2 =
0.024$.
Experimental measurements (\cite{sie73}) extend only to CM energies
greater than 30\keV, for which the $s$-wave resonant contribution is
negligible for $\theta_0^2\lesssim0.5$.  Lower-energy measurements are
therefore needed to determine the nature of the resonance.

\begin{figure}[hbp]
\centering{\epsfig{file=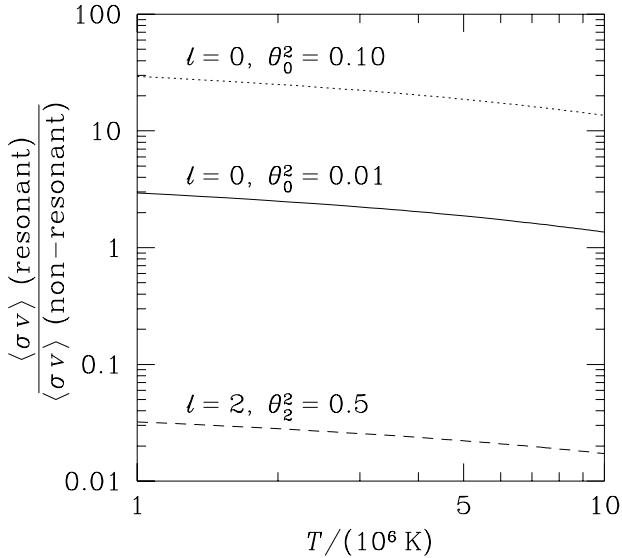,width=\hsize}}
\large\renewcommand{\baselinestretch}{0.7}
\footnotesize
\caption{\protect{\footnotesize Comparison of the thermally averaged
subthreshold resonant rate to the CF88 rate for the reaction \reaction.
We show three cases: (1) angular momentum $\ell=0$ and reduced width
$\theta^2 = 0.01$ ({\em solid line}), (2) $\ell=0$, $\theta^2=0.1$ ({\em
dotted line\/}), (3) and $\ell=2$, $\theta^2=0.5$ ({\em dashed line\/}).
\label{f:ratio}}}
\end{figure}

Upon specifying $\theta_\ell^2$, we compute the resonant cross-section.
Because the reaction proceeds through the tail of the resonance
($E-E_r>\Gamma$), we write $\sigma = S(E) E^{-1}
\exp\left[-(\EG/E)^{1/2}\right]$, where $S(E)$ depends only weakly on
the energy.  Averaging over a Maxwellian velocity distribution then
amounts to evaluating about the peak of the integrand, $E_0\approx(\kB T
\EG^{1/2}/2)^{2/3}$.  We show in Figure \ref{f:ratio} the ratio of the
thermally averaged resonant ($E_r = -25.7\keV$) cross section to that of
CF88, for three different combinations of $\ell$ and $\theta^2$: (1)
$\ell=0$, $\theta_0^2=0.01$ ({\em solid line\/}); (2) $\ell=0$,
$\theta_0^2=0.10$ ({\em dotted line\/}); and (3) $\ell=2$,
$\theta_2^2=0.5$ ({\em dashed line\/}).

As is evident from Figure \ref{f:ratio}, a $d$-wave resonance contributes
little.  An $s$-wave resonance, however, will actually dominate the
\reaction\ cross-section for $\theta_0^2\gtrsim0.01$.  We compute the
astrophysical $S$-factor for this case,
\begin{equation}\label{e:S-factor}
   S(0) = 47.0\left(\frac{\theta_0^2}{0.01}\right)\MeV\barn.
\end{equation}
This is larger than the combined $S$-factor estimated by Sierk \&
Tombrello (1973).  A fit to the \reaction\ rate is
\begin{eqnarray}\label{e:NaSigmaV}
\lefteqn{\left.N_{\rm\!A}\langle\sigma v\rangle\right|_{\ell=0}
= 
  3.53\ee{14}
   \left(\frac{\theta_0^2}{0.01}\right)
   \left(\frac{T}{10^6\K}\right)^{-1.367}} \nonumber\\ 
   && \times\exp\left[-105.326\left(\frac{T}{10^6\K}\right)^{-1/3}\right]
   \cm^3\second^{-1}\gram^{-1}.
\end{eqnarray}
In constructing this fit, we have neglected the nonresonant
contribution to the total cross section; this is permissible for
$\theta_0^2\gtrsim 0.05$.  The total destruction rate for \Be\ is then
the rate for \reaction\ (eq.\ [\ref{e:NaSigmaV}]) added to the rate for
$\Be(p,d)2\,\He$ (CF88).

\section{Observational Constraints on the Cross Section}
\label{s:constraints}

When observations of beryllium in pre--main-sequence stars of mass
$M\lesssim0.3\Msun$ become possible, the reaction rate will be easily
constrained.  For example, a star of mass 0.2\Msun\ will deplete \Be\ at
a considerably (22\%) younger age if the resonance is $s$-wave and
$\theta_0^2=0.1$ (\cite{ush98}).  Observations of lithium abundances
(Basri, Marcy, \& Graham 1996; \cite{bil97}; \cite{ush98}), as well as
fitting to the main-sequence turnoff, can independently constrain the
ages and masses of cluster members.

Stars heavier than $\sim0.3\Msun$ form radiative cores before their
center is hot enough to destroy beryllium.  As the convective zone moves
outward in mass, the base temperature passes through a maximum (e.g.,
\cite{dan94}; \cite{for94}).  As first pointed out by Bodenheimer
(1966), the base of the convective zone at maximum temperature is hot
enough ($T\gtrsim 4\ee{6}\K$) to burn beryllium for a short time during
pre--main-sequence contraction.  The ZAMS beryllium depletion is then
determined by comparing the depletion timescale (\cite{bil97}),
\begin{equation}
   \tdepl\equiv\frac{1}{n_{\rm H}\langle\sigma v\rangle},
\end{equation}
($n_{\rm H}$ is the number density of hydrogen) to the time spent at
maximum temperature by the convective zone (approximately the age of the
star).  Because higher mass stars develop radiative cores at colder
central temperatures, the ZAMS abundance of beryllium increases with
mass.  A star of mass 0.5\Msun\ at an age of 300\Myr\ will be 50\%
depleted (using CF88 rates), while an $0.8\Msun$ star of the same age
will only be 3\% depleted (\cite{for94}).  Standard models of
main-sequence stars more massive than 0.7\Msun\ do not show appreciable
beryllium destruction while on the main sequence.  Hence, a lack of
beryllium in stars more massive than $\sim0.6\Msun$ is an indication
that either the reaction rate is underestimated or that other
mechanisms are at work.  Because the various mixing mechanisms discussed
in \S~\ref{s:motivation} occur on the main sequence, younger
clusters are promising targets for study.  Recent observations of
late-type dwarf stars in the Hyades (\cite{gar95}) are consistent with
no depletion of beryllium for $\Teff>5250\K$.  The age of the Hyades is
$625\pm50\Myr$ (\cite{per98}), so a star of that effective temperature
is on the main sequence and has a mass of about 0.9\Msun\
(\cite{dan94}).  The temperature at the base of the convective zone
($\approx2.6\ee{6}\K$) is far too cold to cause appreciable depletion of
beryllium, even if the resonance were $s$-wave.  At an age of 3\Myr, the
base of the convective zone reaches its maximum temperature of
$4\ee{6}\K$ (\cite{dan94}).  If the depletion timescale is to be longer
than the contraction timescale, $t_d\gg2\Myr$, then we require that
$\theta_0^2\lesssim 0.1$.  A less massive Hyades member that is
beryllium-depleted will place a stronger lower bound on $\theta_0^2$.

\section{Implications}\label{s:implications}

For different combinations of angular momentum and reduced width, we
have estimated the \reaction\ $S$-factor (eq.\ [\ref{e:S-factor}]) and
thermally averaged reaction rate (eq.\ [\ref{e:NaSigmaV}]).  Neither
experimental measurements of the $S$-factor (\S~\ref{s:experiment}) nor
observations of low-mass stars in the Hyades (\S~\ref{s:constraints})
rule out an $s$-wave resonance; at best, the reduced width is only
constrained to be $\theta_0^2\lesssim0.1$.  This reduced width roughly
doubles the standard (CF88) rate at typical stellar temperatures where
beryllium is destroyed.  Because the reaction rate at
$T\approx4\ee{6}\K$ is roughly proportional to $T^{21}$, a doubling of
the rate corresponds to a reduction of 10\% in the temperature at which
beryllium destruction occurs.  The depth of the beryllium preservation
region is therefore reduced.  As we sketched in \S~\ref{s:motivation},
the differences in beryllium and lithium depletion patterns can
potentially discriminate between the many proposed depletion mechanisms.

While experimental measurements at laboratory energies $E_{\rm
lab}<30\keV$ would ideally yield the $S$-factor, stellar observations of
very low mass stars can possibly constrain the reduced width, if not the
entrance channel angular momentum.  There are few other areas of
astrophysics where observations can inform nuclear physics.  Although
the best prospects---fully convective ($M<0.3\Msun$), pre--main-sequence
stars---are too dim in the UV, observations of beryllium-deficient, ZAMS
stars heavier than roughly 0.7\Msun\ would be persuasive evidence for an
enhanced beryllium destruction rate.

\acknowledgements 

Thanks to Lars Bildsten for suggesting this project and for many helpful
discussions, to Richard Boyd for thoughtful comments on the manuscript,
and to Michael Wiescher for providing a tabulated numerical calculation
of the penetration factor
and for helpful comments.  E. F. B. was supported by a NASA GSRP
Graduate Fellowship grant NGT-51662.

\end{document}